\documentclass{elsart}

\usepackage{natbib}

\usepackage{graphicx}
\newcommand{\gr}{$\gamma$-ray \,}

\newcommand{\rxj}{RX~J1713.7-3946 }
\newcommand{\grs}{$\gamma$-rays \,}

\begin{document}

\begin{frontmatter}


\title{Cosmic Ray Acceleration by Supernova Shocks}

\author{E. G. Berezhko}

\ead{berezhko@ikfia.ysn.ru}

\address{Yu. G. Shafer Institute of Cosmophysical Research and Aeronomy\\
31 Lenin Ave., 678980 Yakutsk, Russia}

\begin{abstract}
We analyse the results of recent measurements of nonthermal emission
from individual supernova remnants (SNRs) and their correspondence to
the nonlinear kinetic theory of cosmic ray (CR)
acceleration in SNRs. 
It is shown that the theory fits these
data in a satisfactory way and 
provides the strong evidences for the efficient CR production in SNRs accompanied
by significant magnetic field amplification.
Magnetic field amplification leads to considerable increase of CR maximum energy 
so that the spectrum of
CRs accelerated in SNRs is consistent with the requirements
for the formation of Galactic CR spectrum up to the energy $\sim 10^{17}$~eV. 
\end{abstract}


\end{frontmatter}

\section{Introduction}
The main reason why supernova remnants
(SNRs) are considered as a cosmic ray (CR) 
source is a simple argument about the energy 
required to sustain the Galactic cosmic ray (GCR) 
population against loss by escape, nuclear 
interactions and ionization energy loss. 
The mechanical energy input to the Galaxy from each supernova (SN) is about 
$10^{51}$ erg so that with a rate of about one every 30 years the total mechanical 
power input from supernovae is of the order $10^{42}$ erg/s 
\citep[e.g.][]{berezinetal90}. 
Thus supernovae have 
enough power to drive the GCR acceleration if there exists a 
mechanism for channeling about 10\% of the mechanical energy into relativistic 
particles. The high velocity ejecta produced in the supernova explosion 
interacts with the ambient medium to produce a system of strong shocks. The 
shocks in turn can pick up a few particles from the plasma flowing into the 
shock fronts and accelerate them
to high energies.

The only theory of particle acceleration which at present is
sufficiently well developed and specific to allow quantitative model
calculations is
diffusive acceleration applied to the strong shocks associated with
SNRs \citep[e.g. see][for review]{drury83, blei87, bk88,jel91,md01}. 
Considerable efforts have been made during the last years to empirically
confirm the theoretical expectation that the main part of GCRs indeed
originates in SNRs.Theoretically
progress in the solution of this problem has been due to the development
of the kinetic nonlinear theory of diffusive shock acceleration 
\citep{byk96, bv97, bv00a}.
The theory includes all the most relevant physical
factors, essential for SNR evolution and CR acceleration, 
and it is able to make quantitative predictions of the expected properties
of CRs produced in SNRs and their nonthermal radiation. 

Detail information about high-energy CR population in SNRs can be
obtained from the observations of the nonthermal 
emission produced by accelerated CRs
in SNRs. The electron CR component is evident
in a wide wave length range by the radiation that they produce
in SNRs,  from radio to $\gamma$-ray emission, whereas in the case of the
nuclear CR
component a $\gamma$-ray detection is the only possibility to find it. 
If this nuclear component is
strongly enhanced inside SNRs then through inelastic nuclear collisions,
leading to pion production and subsequent decay, $\gamma$-rays will be
produced at the detectable level.

High-energy $\gamma$-rays can also be produced by CR electrons due to the
inverse Compton (IC) scattering of the background photons. Therefore it is not
so obvious which CR component (nuclear or electron) produces 
TeV emission, detected from
Cassiopeia~A (Cas~A) \citep{aha01}, 
RXJ1713.7-3946 \citep{mur00,enomoto02,aha05,aha06a} and Vela~Jr \citep{aha06b}. 
For this determination the SNR magnetic field value plays
the crucial role. 
The relative role of electrons in $\gamma$-ray production is low if SNR
magnetic field is strongly amplified. In addition,
if the magnetic fields in SNRs are considerably larger than
the typical interstellar value, then SNRs are able to generate a
power-law spectrum of accelerated CRs
at least up to the knee energy \citep{bk99,bv04}.
In fact, there is strong evidence that the actual magnetic field in SNRs
is indeed intrinsically enhanced:
the analysis of the nonthermal emission of young SNRs show
that all of them have
amplified fields \citep{vbk05, pariz06}.
Required strength of the magnetic field, that
is significantly higher than the 
interstellar medium (ISM) field, has to be attributed to nonlinear
field amplification at the SN shock by CR acceleration itself, in fact by
the dominant nuclear CR component \citep{lb00,bell04}.

The application of kinetic
theory to individual SNRs \citep[see][for a review]{ber05} has demonstrated its
power in explaining the observed SNR properties and in predicting new effects
like the extent of magnetic field amplification, leading to the concentration
of the highest-energy electrons in a very thin shell just behind the shock.
The theory also predicted the enhancement of the high-energy galactic \gr
background radiation due to the contribution of CRs confined in parent SNRs,
that was partly confirmed by the measured background TeV-emission.

In order to perform the detail comparison of theoretical expectation with the
experiment one needs sufficient number 
of individual SNRs with known values of relevant
physical parameters such as the age, distance, ISM density.
Unfortunately the number of such SNRs are very limited, especially thous which
are seen in all wavelength from radio to \gr. It is why
every new experimentally established property of nonthermal emission of
SNR represents considerable interest for theoretical
analysis. Therefore we start this review with a brief consideration of three
SNRs for which essentially new information about the efficiency of
CR production was recently obtained. Then we shall
consider the correspondence of the interior magnetic field values extracted
from the analysis of the experimental data with theoretical expectation.
Finally, we discuss the maximum energy of CRs which can be achieved during their
acceleration in SNRs, taking into account realistic values of magnetic field in
SNRs.

\section{Evidences for efficient CR production in SNRs}
Experimental and theoretical study of young galactic SNRs, Tycho's SNR, \rxj and
SN~1987A, performed recently provides very useful information about CR
acceleration by SN shocks. Therefore below we briefly analyse these results.

\subsection{Tycho's supernova remnant}
The kinetic
theory has been applied in detail to Tycho's SNR, the result of a type
Ia SN explosion in a (roughly) uniform interstellar medium
(ISM), in order to compare
the results with existing data, using a stellar ejecta mass
$M_{ej}=1.4M_{\odot}$, distance $d=2.3$~kpc, and ISM number density $N_H=0.5$~
H-atoms cm$^{-3}$ \citep{vbkr,vbk05}. For these parameters a total hydrodynamic
explosion energy $E_{sn}=0.27\times 10^{51}$~erg was derived to fit the
observed size $R_s$ and expansion speed $V_s$. A rather high downstream
magnetic field strength $B_d\approx 300$~$\mu$G is 
needed to reproduce the observed steep and concave
radio spectrum \citep[see also][]{rel92}
and to ensure a smooth cutoff of the synchrotron emission in the
X-ray region. Overall, very good
consistency of the predictions of the nonlinear theory with the existing
observational data was achieved.

Using Chandra X-ray observations \citet{warren05} have recently estimated
the ratio between the radius $R_\mathrm{c}$ of the contact discontinuity (CD),
separating the swept-up ISM and the ejecta material, and the radius
$R_\mathrm{s}$ of the forward shock. The large mean value
$R_\mathrm{c}/R_\mathrm{s}=0.93$ of this ratio was interpreted as evidence for
efficient CR acceleration, which makes the medium between those two
discontinuities more compressible. This result was analysed within already
performed kinetic nonlinear approach \citep{vbk06}.
Note, that the increase of the ratio $R_\mathrm{c}/R_\mathrm{s}$ due to the SN
shock modification by efficiently accelerated CRs was predicted for the case of
Kepler's SNR by \citet{dec00}.

\begin{figure}
\centering
\includegraphics[width=7.5cm]{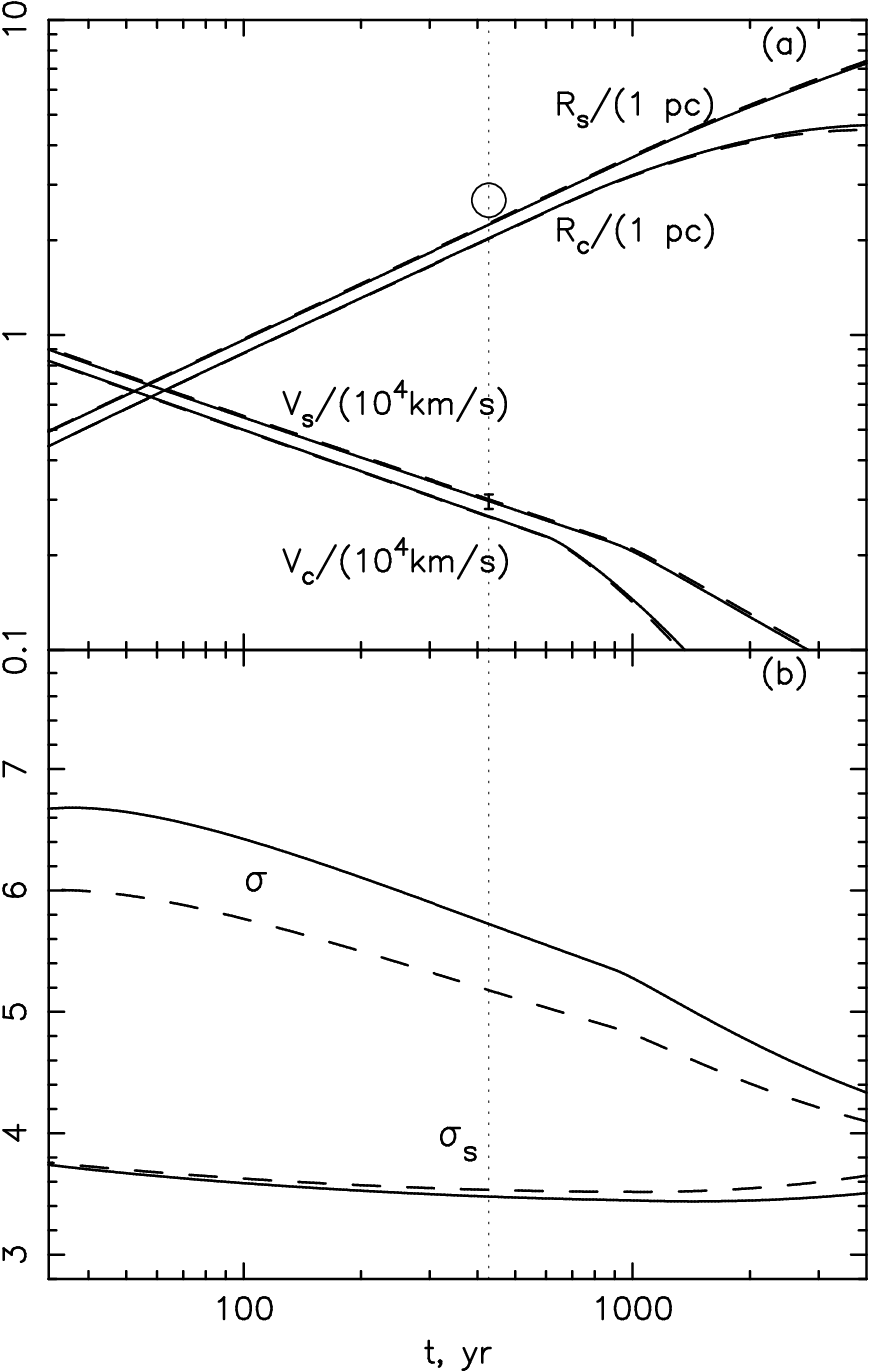}
\caption{(a) Shock radius $R_\mathrm{s}$, contact
discontinuity radius $R_\mathrm{c}$, shock speed $V_\mathrm{s}$, and contact
discontinuity speed $V_\mathrm{c}$, for Tycho`s SNR as functions of time,
including particle acceleration; 
(b) total shock ($\sigma$) and subshock
($\sigma_\mathrm{s}$) compression ratios \citep{vbk06}. The {\it dotted vertical line} marks
the current epoch. The {\it solid and dashed lines} correspond to the internal
magnetic field strength $B_\mathrm{d}=240$~$\mu$G and
($B_\mathrm{d}=360$~$\mu$G), respectively.  The observed mean size and speed of
the shock, as determined by radio measurements
\citep{tg85}, are shown as well.} 
\label{f1}
\end{figure}

\begin{figure}
\centering
\includegraphics[width=7.5cm]{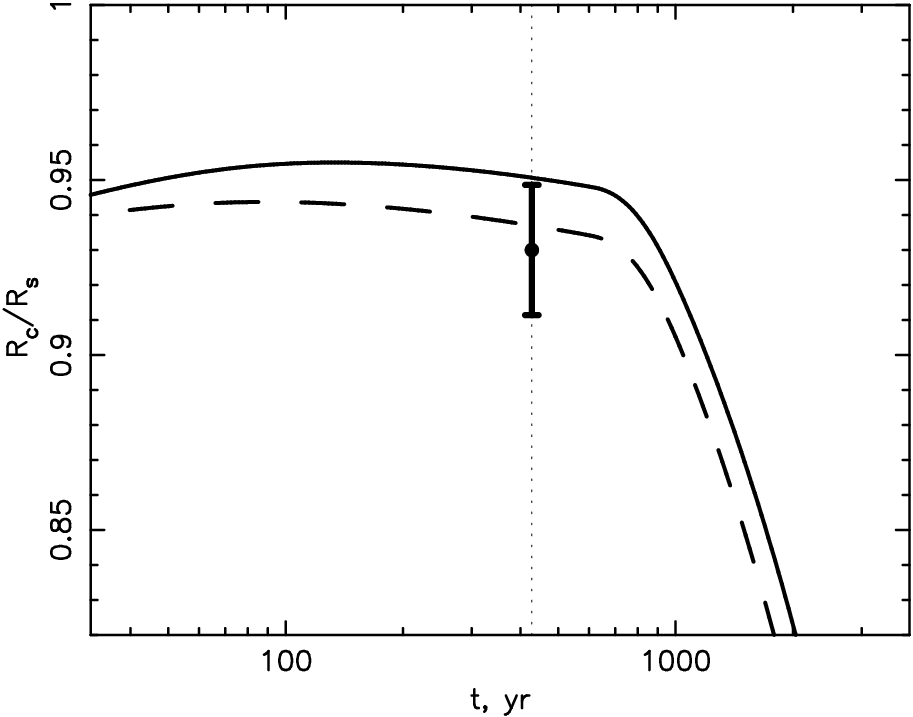}
\caption{
The ratio $R_\mathrm{c}/R_\mathrm{s}$ of the radii of the contact
discontinuity and the forward shock
of Tycho's SNR as a function of time \citep{vbk06}. Solid and
dashed lines correspond to the same two cases as in Fig.\ref{f1}. 
The experimental point is taken from \citet{warren05}.}
\label{f2}
\end{figure}

Fig.\ref{f1} and Fig.\ref{f2} show the calculations of shock and CD
related quantities which were part of the earlier considerations
\citep{vbkr,vbk05}, together with the azimuthally averaged experimental data
available at the time. The calculated shock as well as CD radii and speeds are
shown as a function of time for the two different cases of interior magnetic
field strengths $B_\mathrm{d}=240$~$\mu$G and $B_\mathrm{d}=360$~$\mu$G
required by the value of the synchrotron frequency, where the
spectrum hardens relative to the low-energy radio spectrum.

According to Fig.\ref{f1}a Tycho's SNR is nearing the adiabatic phase.
The softness of the observed low-energy radio spectrum
-- relative to a test particle spectrum -- required a
proton injection rate $\eta=3\times 10^{-4}$. This
implies a significant nonlinear modification of the shock at the current age
of $t=428$~yrs. A larger magnetic field lowers the Alfv\'enic Mach number and
therefore leads to a decrease of the shock compression ratio, as seen in
Fig.\ref{f1}b. The result is a total compression ratio $\sigma=5.7$ and a subshock
compression ratio $\sigma_s=3.5$ for $B_\mathrm{d}=240$~$\mu$G. In turn
$\sigma=5.2$, $\sigma_s=3.6$, for $B_\mathrm{d}=360$~$\mu$G.

Therefore, including CR acceleration at the outer
blast wave, the calculated value of the ratio $R_\mathrm{c}/R_\mathrm{s}$ for
$B_\mathrm{d}=360$~$\mu$G is slightly lower than for $B_\mathrm{d}=240$~$\mu$G.
At the current epoch we have $R_\mathrm{c}/R_\mathrm{s}\approx 0.90$ which is
somewhat lower than the value $R_\mathrm{c}/R_\mathrm{s}=0.93$ inferred
from the Chandra observations. 

Projecting a highly structured shell onto the plane of the sky tends to favor
protruding parts of the shell. Therefore the average radius measured in
projection is an overestimate of the true average radius. Analyzing the amount
of bias from the projection for the shock and CD radii \citet{warren05} found a
corrected "true" value $R_\mathrm{c}/R_\mathrm{s}=0.93$ which is lower than
their measured ``projected average'' value $R_\mathrm{c}/R_\mathrm{s}=0.96$.
One needs
to make such a purely geometrical correction independently of the nature of the
factors which produce the shell structures, if one measures the size of the
quasispherical structured shell in projection instead of dealing with a
2-dimensional shell crossection. In the latter case there would be no need for
such a correction.

Irrespective of this need to correct the observations of the CD position, if
one starts from a spherically symmetric calculation of the CD radius as we do,
one has to take into account that the actual CD is subject to the
Rayleigh-Taylor (R-T) instability
\citep[e.g.][]{chevetal92,wang01}, and thus a correction is
needed to compare such a 1-D calculation with observations of the CD. In the
nonlinear regime the instability leads to effective mixing of the ejecta and
swept-up ISM material with ``fingers'' of the ejecta on top of this mixing
region. The latter has a radius larger than $R_\mathrm{c}^\mathrm{1D}$.  The
minimum correction to $R_\mathrm{c}^\mathrm{1D}$ is the radial extent $\Delta
R_\mathrm{c}$ of the mixing region above $R_\mathrm{c}^\mathrm{1D}$. We
estimate $\Delta R_\mathrm{c} \approx 0.5l$, where $l\approx
0.1R_\mathrm{c}^\mathrm{1D}$ is the longest finger size according to the
numerical modeling of \citet{wang01} (albeit without particle acceleration).
This leads to a rough estimate of the corrected CD radius $R'_c = 1.05
R_\mathrm{c}^\mathrm{1D}$.

As it is seen in Fig.2, 
the comparison of the corrected values $R'_\mathrm{c}/R_\mathrm{s}$, according
to the earlier calculations \citep{vbkr,vbk05}, with this experimentally
estimated value $R_\mathrm{c}/R_\mathrm{s}=0.93$ shows quite good
agreement.

The agreement between theoretical solutions for effective
particle acceleration with the measurements of the 
shock and discontinuity radii 
has to be considered as new evidence for strong nonthermal effects in Tycho's
SNR. New
Northern Hemisphere TeV detectors is expected to 
detect this source at TeV-energies in,
predominantly, hadronic $\gamma$-rays at an energy flux level above $2 \times
10^{-13}\mathrm{erg}~\mathrm{cm}^{-2}~\mathrm{s}^{-1}$.  
As a corollary the detection of a TeV signal is not only important
by itself, but it is also crucial for the correct determination of all other
key SNR parameters.

\subsection{Supernova remnant RX~J1713-3946}
RX~J1713-3946 is a shell-type supernova remnant (SNR), located in the Galactic
plane, that was discovered in X-rays with ROSAT \citep{pfef96}.
Further study of this SNR with the ASCA satellite by
\citet{koyama97} and later by \citet{slane99} have shown that the
observable X-ray emission is entirely non-thermal, and this property was
confirmed in later XMM-Newton observations \citep{cassam04}.
The radio
emission of this SNR is weak: only part of the shell could be 
detected in
radio synchrotron emission up to now, with a poorly known spectral form
\citep{laz04}.

RX~J1713-3946 was also detected in very high energy \grs with the
CANGAROO \citep{mur00,enomoto02} and H.E.S.S.
\citep{aha06a} telescopes. Especially the latter observations
show a clear shell structure at TeV energies which correlates well with the
ASCA contours.

The difficulty for the theoretical description is 
the fact that several key parameters of this 
source are either not known or poorly constrained. This already concerns 
the distance and age of the object. 
It was demonstrated that consistent description
of this object is achieved 
\citep{bv06} following present consensus which puts 
the distance at 1 kpc, the age to about 1600 years 
and that the primary explosion must have been a type 
II/Ib SN event with a massive progenitor star whose mass loss in the main 
sequence phase created a hot wind bubble in a high-density environment. 
The solution for the overall remnant dynamics then yields the value for 
the expansion velocity of the outer shock, given the total mechanical 
energy $E_{\mathrm{sn}}=1.8 \times 10^{51}$~erg 
released in the explosion. To obtain a consistent 
solution for the broadband nonthermal emission the injection rate $\eta=3\times 10^{-4}$
and
interior magnetic field $B_\mathrm{d}\approx 130~\mu$G are needed.

\begin{figure}
\centering
\includegraphics[width=13.5cm]{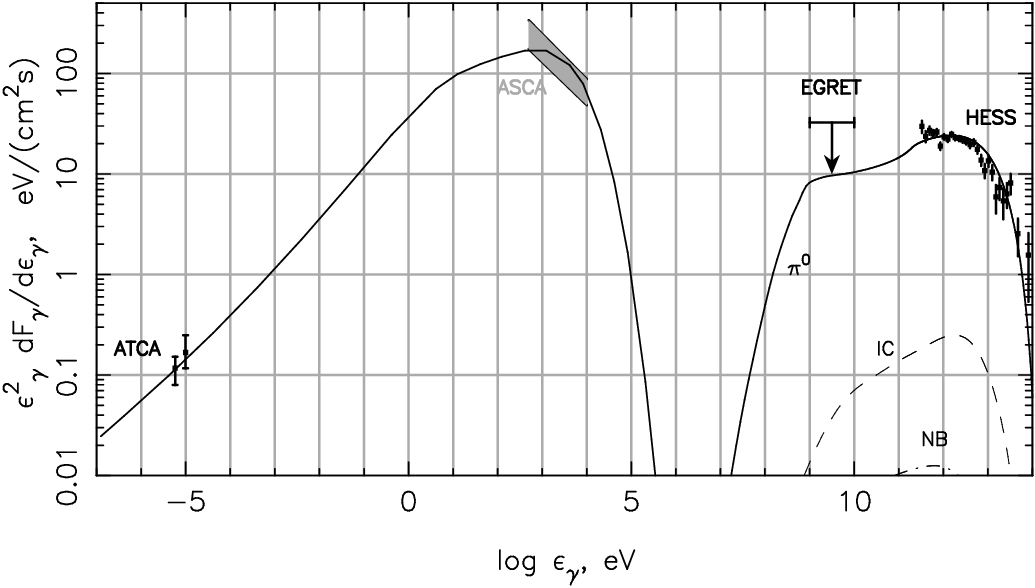}
\caption{Spatially integrated spectral energy distribution of \rxj
\citep{bv06}. The 
{ATCA} radio data \citep[cf.][]{aha06a}, {ASCA} X-ray data
\citep[cf.][]{aha05}, and
{H.E.S.S.} data \citep{aha07} are
shown. The {EGRET} upper limit for the \rxj position \citep{aha05}
is shown as well. The solid curve at
energies above $10^7$~eV corresponds to $\pi^0$-decay \gr emission, whereas the
dashed and dash-dotted curves indicate the inverse Compton (IC) and
Nonthermal Bremsstrahlung (NB) emissions, respectively.
}
\label{f3}
\end{figure}

The calculated overall broadband spectral energy distribution is displayed in
Fig.\ref{f3}, together with the experimental data from {ATCA} at radio
wavelengths, as estimated for the full remnant by \citet{aha06a},
the X-ray data from {ASCA}, 
and the most recently measured TeV gamma-ray spectrum from H.E.S.S.
\citep{aha07}.
The overall fit is impressive, noting that the choice of a few
key parameters like $\eta$, $B_\mathrm{d}$, $E_{\mathrm{sn}}$ 
in the theory allows a spectrum determination over 
more than 19 decades. 
Note, that calculated gamma-ray stectrum 
(Berezhko and V\"olk, 2006) agrees with
the new H.E.S.S. data (Aharonian et al., 2007) even better than
with earlier H.E.S.S. measurements (Aharonian et al., 2006a).

The properties of small scale structures of SNR \rxj seen in X-rays by
\citet{uchiyama03}, and in particular by \citet{hiraga05}, provide even 
stronger evidence that the magnetic field inside the SNR is indeed considerably
amplified.

\begin{figure}
\centering
\includegraphics[width=7.5cm]{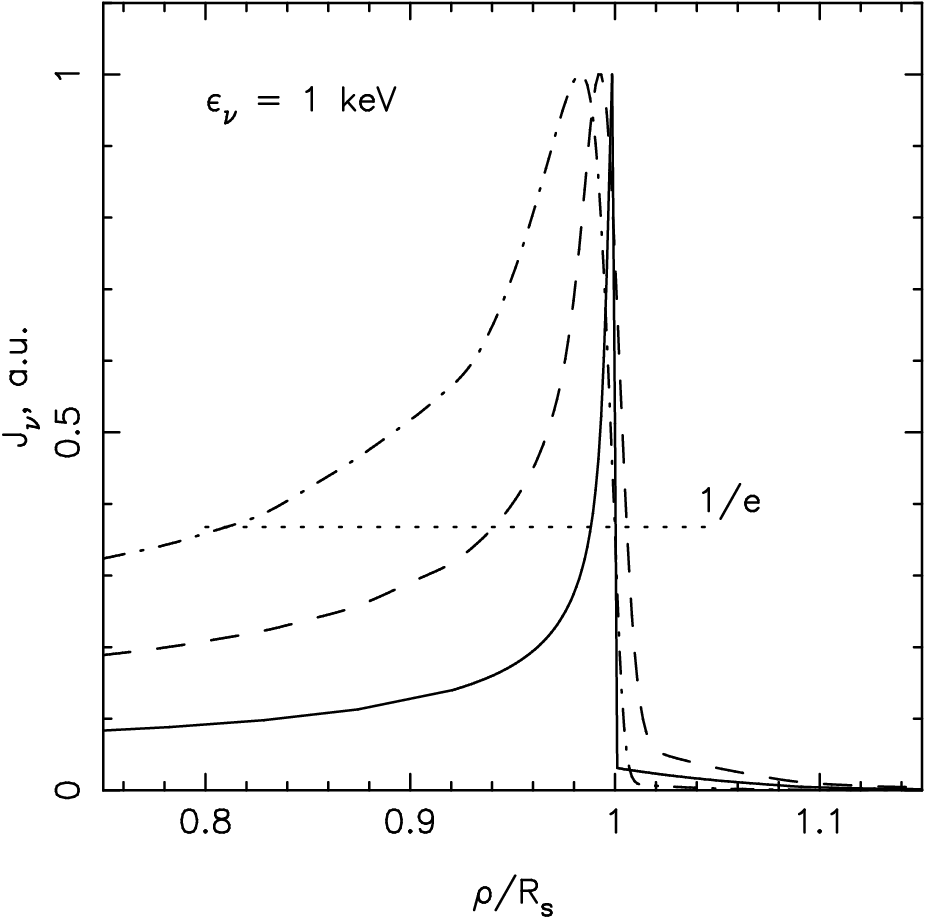}
\caption{Projected radial profile of the X-ray synchrotron emission 
for the energy $\epsilon_\mathrm{\nu}=1$~keV from \rxj
\citep{bv06}. The solid line corresponds
to the high-injection model; the dashed line again represents the above
profile, but smoothed to the resolution of the {XMM-Newton} data used in
\citet{hiraga05}; the dash-dotted line corresponds to the
test-particle limit. 
}
\label{f4}
\end{figure}

In order to find out if the filamentary structure found
by \citep{hiraga05} is indeed consistent with the efficient CR production
projected radial profile $J(\epsilon_{\nu},\rho)$
calculated for the X-ray energy 
$\epsilon_{\nu}=1$~keV is shown in Fig.\ref{f4}. 
The theory predicts the peak of the emission just behind the shock
front with projected
thickness $\Delta \rho/R_\mathrm{s}\approx 10^{-2}$ that 
corresponds to an
angular width $\Delta \psi\approx 0.4'$. This width is significantly thinner
than the observed width of $\Delta \psi\approx 2.2'$. 
However, one should take into
account the smoothing procedure with which the data were
obtained/presented. If we apply a point spread
function of width $\sigma_{\psi}=12.8''$, corresponding to the smoothed XMM
data, to the calculated profile, we obtain the broadened profile , which is
shown in Fig.\ref{f4} by the dashed curve.  It has a width $\Delta 
\rho/R_\mathrm{s}\approx 6.6
\times 10^{-2}$, or $\Delta \psi\approx 2'$, and is consistent with
the observational profile of \citet{hiraga05}.

In addition the projected radial profile (dash-dotted 
line) which corresponds to the test particle limit
is presented in Fig.\ref{f4}. 
In this case the proton injection rate was taken artificially so small
$(\eta=10^{-5})$, that accelerated CRs do not produce any significant nonlinear
effect, like shock modification and/or magnetic field amplification.
Therefore interior magnetic field value $B_\mathrm{d}=20$~$\mu$G
is much lower in this case. Due to this fact
synchrotron losses are not relevant on account of the low magnetic 
field, the peak of the emission is broader 
by a factor of five, inconsistent with the {XMM} observations.

\begin{figure}
\centering
\includegraphics[width=7.5cm]{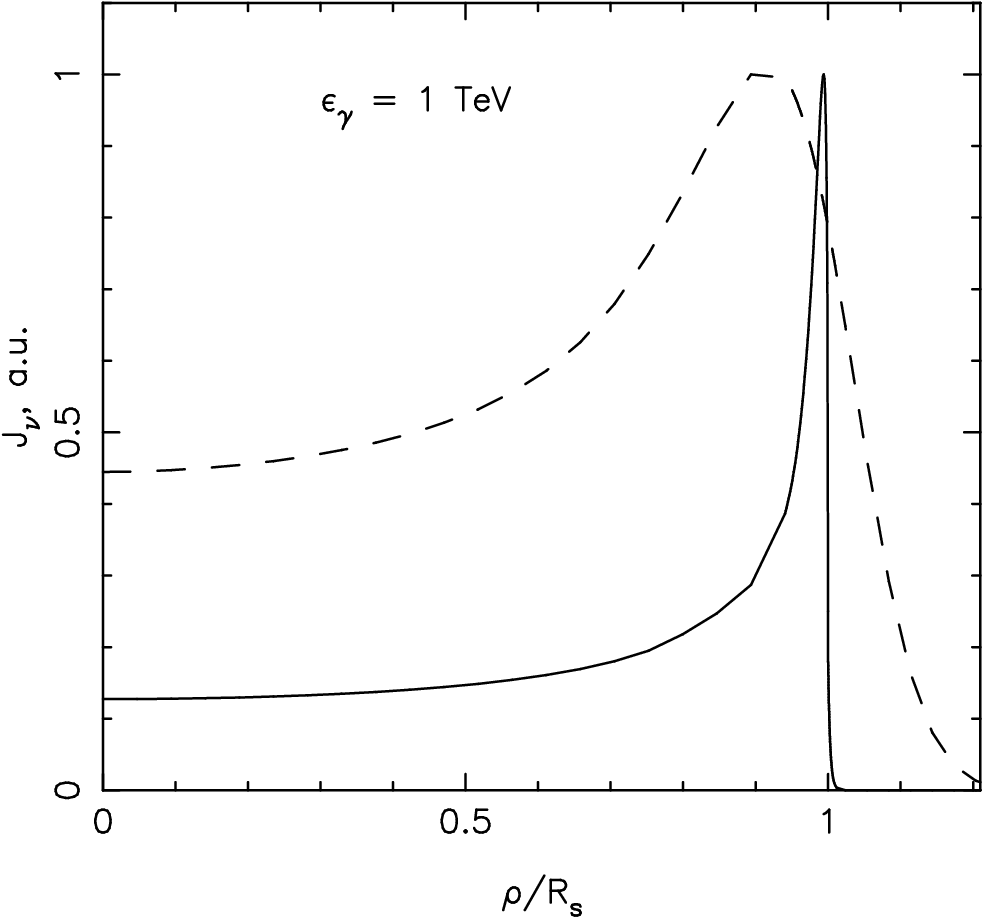}
\caption{The projected radial profile of \gr 
emission for \gr energies $\epsilon_\mathrm{\gamma}=1$~TeV \citep{bv06}.  
The calculated radial profile is
 given by the solid line; the dashed line represents the calculated
 profile smoothed with a Gaussian point spread function of the width
 $\sigma_{\psi}=0.05^\circ$. For purposes of presentation both profiles are
 normalised to their peak values in this figure.
}
\label{f5}
\end{figure}

The projected radial $\gamma$-ray emission profile, calculated for the energy
$\epsilon_\mathrm{\gamma}=1$~TeV, is presented in Fig.\ref{f5}. 
As a result of the
large radial gradient of the gas density and the CR distribution inside the
SNR, the theoretically predicted radial profile of the TeV-emission is
concentrated within a thin shell of width $\Delta \rho\approx 
0.1R_\mathrm{s}$. Since the H.E.S.S. instrument has a finite 
angular resolution we
present in Fig.\ref{f4} also the modified radial profile smoothed with the Gaussian
point spread function with $\sigma_{\rho}=\Delta \rho= 0.08R_\mathrm{s}$, 
that
corresponds to the angular resolution $2\sigma_{ \psi}=0.1^ \circ$.

As shown by Fig.\ref{f5} the smoothed radial profile of the TeV-emission is much 
broader and -- what is most interesting -- it is characterised by a 
maximum to minimum intensity ratio 
$J^{\mathrm{max}}_{\gamma}/J^{\mathrm{min}}_{\gamma} =2.2$. Such a ratio 
is consistent with the H.E.S.S. measurement \citep{aha06a}
which obviously gives only a lower limit to the sharpness of the 
$\gamma$-ray profile. Note that the broad radial profile of the 
TeV-emission with $J^{\mathrm{max}}_{\gamma}/J^{\mathrm{min}}_{\gamma} 
\approx 2$ measured by the H.E.S.S. instrument is indirect evidence 
that the actual radial profile is significantly sharper, with a higher 
ratio $J^{\mathrm{max}}_{\gamma}/J^{\mathrm{min}}_{\gamma} > 2.2$. 

We conclude that the present observational knowledge of SNR \rxj can be
interpreted by a source which ultimately converts about 10\% of the mechanical
explosion energy into nuclear CRs and that the observed high energy \gr
emission of SNR \rxj is of hadronic origin. Note, that \citet{malkov05} have
obtained the same type of conclusion suggesting a rather different \gr spectrum
formation scenario, compared with \citet{bv06}.

\subsection{Supernova remnant SN~1987A}
Supernova 1987A occurred in the Large Magellanic Cloud.
It has been extensively studied in all wavelengths from radio to \gr.
The initial short outburst of radio emission \citep{turtle87} is attributed
to the synchrotron emission of electrons accelerated by the SN shock propagated
in the free wind of
presupernova star, which was the blue supergiant (BSG) 
\citep{chf87}. After about 3 years
radio emission was detected  again \citep{ss92, gae97}
as well as the monotonically increased X-ray emission \citep{gro94, has96}.
This second increase of emission is attributed to
the entrance of the outer SN shock into the thermalized BSG wind and then
in the H~II
region occupied by much more dense matter, consists of the 
swept up wind of red
supergiant (RSG) progenitor star \citep{chd95}.

The advantage of SN~1987A for theoretical consideration is that
the key parameter values are very well determined:
stellar ejecta mass
$M_{ej}=10M_{\odot}$, distance $d=50$~kpc, 
hydrodynamic explosion energy $E_{sn}=1.5\times 10^{51}$~erg
\citep[e.g.][]{mccray93}.
During an initial period the shell material has a broad distribution in
velocity $v$. The fastest part of this ejecta distribution is described by
a power law $dM_{ej}/dv\propto v^{2-k}$
with $k=8.6$.

The kinetic nonlinear model for CR acceleration in SNRs has been in
detail applied to SN~1987A, 
in order to compare its results with 
observational properties \citep{bk06}.

A rather high downstream magnetic field strength
$B_d> 1$~mG is
needed to reproduce the observed steep radio spectrum
\citep{bk00}. The required strength of the magnetic field
have to be attributed to
nonlinear field amplification at the SN shock by CR
acceleration itself.
Since for all the
thoroughly studied young SNRs, the ratio of magnetic field
energy density $B_0^2/8\pi$ in the upstream region of the shock precursor to
the CR pressure $P_c$ is about 
$B_0^2/(8\pi P_c) \approx 5\times 10^{-3}$ \citep{vbk05} and
CR pressure in young SNRs has a typical value $P_c\approx 0.5 \rho_0 V_s^2$,
the upstream magnetic field $B_0=B_d/\sigma$ was taken in the form
\begin{equation}
B_0=\sqrt{2\pi \times 10^{-3}\rho_0 V_s^2}.
\end{equation}

\begin{figure}
\centering
\includegraphics[width=6.5cm]{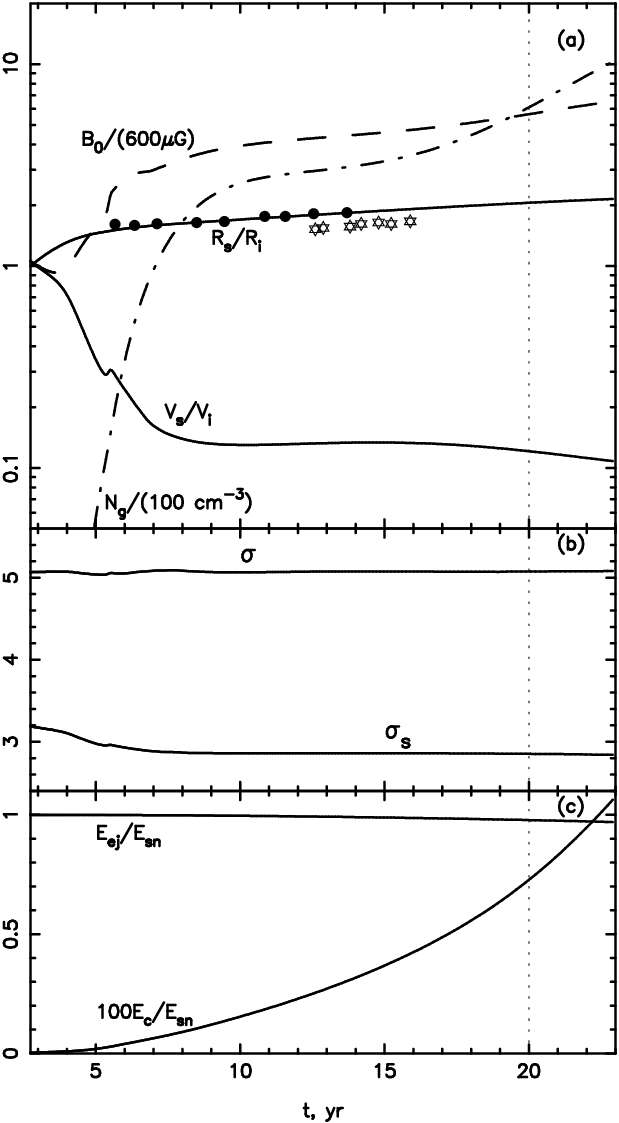}
\caption{(a) Shock radius $R_\mathrm{s}$, 
shock speed $V_\mathrm{s}$, 
gas density $N_g$ and upstream magnetic field $B_0$ at the current shock
position; 
(b) total
shock ($\sigma$) and subshock ($\sigma_\mathrm{s}$) compression ratios;
(c) kinetic energy of ejecta $E_{ej}$ and accelerated CR energy content $E_c$
as functions of time for SN~1987A
\citep{bk06}. The observed radius of
the SN shock, as determined by radio \citep{manch02} 
and X-ray measurements \citep{park04}, are shown by circles and stars
respectively. The scaling values are 
$R_i=R_T=3.1\times 10^{17}$~cm and $V_i=28000$~km/s.
The {\it dotted
vertical line} marks the current epoch. 
\label{f6}}
\end{figure}

Calculated shock radius $R_s$ and speed 
$V_s$ shown in Fig.\ref{f6}a as a function of
time are in satisfactory agreement with the values obtained on the basis of
radio and X-ray measurements. Note that radio data 
compared with X-ray data gives larger shock size
at any given time $t$ and calculated radius 
$R_s(t)$ goes between these two sets of data.

To fit the spectral shape of the observed
radio emission a proton injection rate
$\eta=3\times 10^{-3}$ is needed. This leads to a significant nonlinear modification of
the shock: as it is seen in Fig.\ref{f6}b total shock compression ratio
$\sigma\approx 5.3$ is essentially larger
and a subshock compression ratio $\sigma_s\approx 2.8$ is lower 
than classical value 4.

Since during the last 15~yrs the shock speed 
and the gas density are almost constant CR
energy content grows roughly linearly with time (see Fig.\ref{f6}c). It gives the natural
explanation of the linear increase of the radio emission detected during this
evolutionary period.

\begin{figure}[t]
\centering
\includegraphics[width=7.5cm]{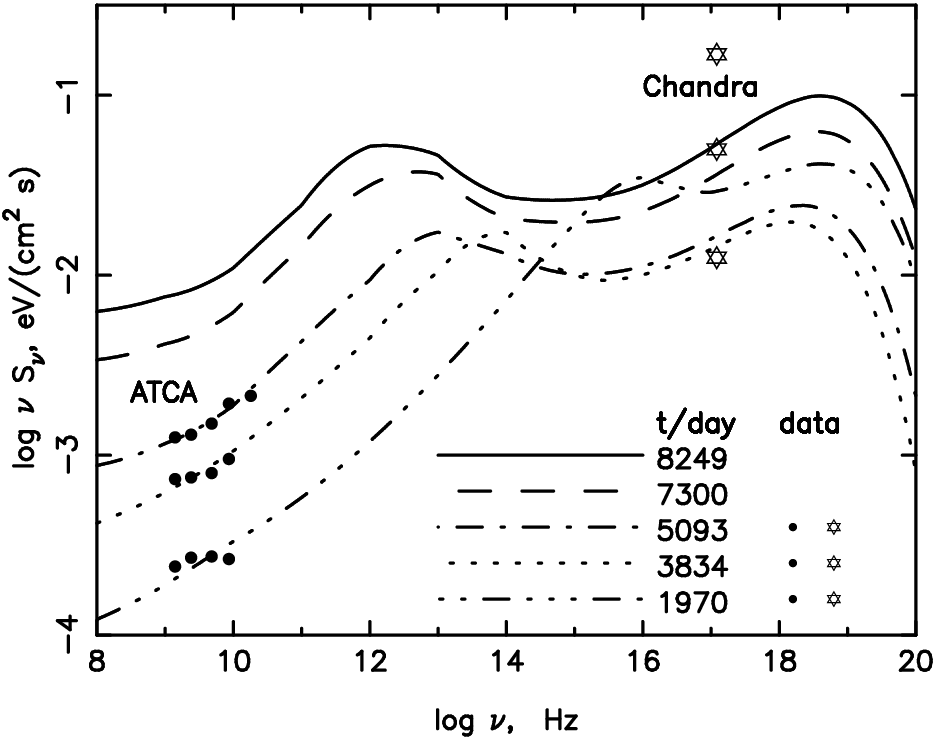}
\caption{
Synchrotron energy spectrum of SN~1987A, calculated for 
the five evolutionary epochs \cite{bk06}. The 
ATCA radio \citep{manch02, mg01}
and Chandra X-ray \citep{park04} data for three epochs are shown as
well. Higher measured fluxes correspond to later epoch.
}
\label{f7}
\end{figure}

Strongly modified SN shock generates CR spectrum 
$N\propto p^{-\gamma}$, which is very steep at momenta
$p<m_pc$, with index $\gamma =(\sigma_s+2)/(\sigma_s-1)\approx 2.7$.
CR electrons with such a spectrum produces synchrotron radioemission
spectrum $S_{\nu}\propto \nu^{-\alpha}$ with spectral index $\alpha
=(\gamma-1)/2\approx 0.9$, that very well corresponds to the experiment,
as it is seen in Fig.\ref{f7}, where synchrotron energy spectra
$\nu S_{\nu}$,
calculated for five subsequent epoch together with the experimental data are
presented.
Note that CR spectrum has a concave shape: it becomes flatter at higher momenta
$p$.
As a consequence synchrotron spectrum $S_{\nu}(\nu)$ is also concave as it is
clearly seen in Fig.\ref{f7} at $\nu<10^{12}$~Hz. Radio data reveal this property in
good consistency with theoretical prediction.

Strong downstream magnetic field $B_d\approx 15$~mG, that corresponds to the
upstream field $B_0\approx 3$~mG (see Fig.\ref{f7}), provides synchrotron cooling
of electrons with momenta $p>10m_pc$ that in turn makes 
synchrotron spectrum at high frequencies $\nu>10^{12}$~Hz
very steep (see Fig.\ref{f7}). 
Concave shape of electrons continuously produced at the shock front together with
their synchrotron cooling lead to a formation of two peaks in synchrotron energy
spectrum $\nu S_{\nu}$. The first one at $\nu \approx 10^{12}$~Hz corresponds to
CR electron momentum $p\approx 10m_pc$ above which synchrotron energy looses
are relevant, whereas the second peak at $\nu \approx 10^{18}$~Hz corresponds
to the maximum momentum $p\approx 10^4m_pc$ of accelerated electrons.
Under this condition calculated
synchrotron flux at frequency $\nu\approx 10^{17}$~Hz, which
corresponds to the photon energy $\epsilon_{\gamma}=0.5$~keV, 
is below the measured
flux at the epochs $t>3000$~d.
Since the contribution of the nonthermal radiation in the
observed X-ray emission of SN~1987A is not 
very well known \citep[e.g.][]{mich02},
the observed X-ray flux has to be considered as 
the upper limit for the expected nonthermal emission. 
At early epoch $t<2500$~d however the calculated flux
exceeds the measured one (see the curve, corresponding $t=1970$~d in
Fig.\ref{f7}).
This can be considered as indication, that the actual magnetic field
$B_0$ is few times larger then given by the Eq.(1).

At lower magnetic field $B_d \approx 2$~mG 
synchrotron losses of high energy CR electrons
are considerably smaller compared with the previous case. Due to this fact
synchrotron spectra considerably exceeds at any given epoch the measured Chandra
flux \citep{bk06}. Therefore one can conclude, that the actual interior magnetic field
strength is not lower than $5$~mG.

We see that quite a reasonable 
consistency of kinetic nonlinear theory with most of the observational
data of SN~1987A is achieved, that gives the evidence of the
efficient CR production leading to a strong shock modification and 
strong magnetic field amplification. 

Calculated \gr integral energy flux  
at all energies is dominated by the $\pi^0$-decay component \citep{bk06}.
At the current epoch the expected \gr energy flux at
TeV-energies is about 
$\epsilon_{\gamma}F_{\gamma}\approx 2\times 10^{-13}$~erg/(cm$^2$s) 
and during the next four years it is expected to grow by a factor of two. 
The detection of high energy \grs from SN~1987A, which could be done by
instruments like H.E.S.S., would provide very clear evidence for the efficient
acceleration of nuclear CRs.

\subsection{Discussion}
Detailed consideration performed within a frame of nonlinear
kinetic theory demonstrates
that the CR production efficiency in young SNRs is consistent with requirements
for the Galactic CR sources. Recent experimental and theoretical studies of
Tycho's SNR, \rxj and SN~1987A provides new very strong evidence for that.

Consideration performed by \citet{hillas05} led him to the opposite conclusion:
that TeV \gr emission from the mostly fairly young SNRs (SN~1006, Cas~A and
Tycho's) is unexpectedly low. Note, however, that this consideration is based on
the simplified estimates, which does not take into account such 
an important aspect
as the necessity of renormalization of CR production efficiency, predicted in
spherical approach.
We have argued before that ion injection is quite efficient at the 
quasiparallel portions of the shock surface and it 
is expected to be strongly suppressed at the 
quasiperpendicular part of the shock \citep[see][for details]{vbk03}.
Therefore one should renormalize the results 
for the nucleonic spectrum, calculated within the spherically symmetric 
model. The lack of symmetry in the actual SNR can be approximately taken 
into account by a renormalization factor $f_{\mathrm{re}}<1$, roughly 
$f_{\mathrm{re}}=0.15$~to $0.25$, which diminishes the nucleonic CR 
production efficiency, calculated in the spherical model, and all effects 
associated with it.

For SN~1006 which appears to approximate this ideal case, efficient CR
production is expected to arise within two polar regions, where the SN
shock is quasi-parallel. The calculated size of the efficient CR production regions
which amounts to about 20\% of the shock surface 
corresponds very well to
the observed sizes of the bright X-ray synchrotron emission regions \citep{vbk03}. 
This implies a renormalization factor of 0.2.
In addition, the distance to the object
$d=1.1$~kpc used by \citet{hillas05} for the case of
SN~1006 is considerably
lower than the most reliable determination of the distance $d=2.2$~kpc
to SN~1006 \citep{winkler03}. 

Due to the above two factors the estimate of TeV \gr flux from SN~1006, made by
\citet{hillas05}, should be reduced by a factor of 20, so that it will be in
agreement with H.E.S.S. upper limit.
Note, that the study of SN~1006 on the basis of kinetic nonlinear theory
demonstrated that the lack of a TeV signal 
that follows from the non-detection by the H.E.S.S. instrument does not 
invalidate the theoretical picture which gives a consistent description of the 
nonthermal emission characteristics; it rather implies a constraint on the 
ambient gas density $N_\mathrm{H} < 0.1$~cm$^{-3}$ \citep{kbv05}.
Using XMM-Newton
X-ray observations \citet{acero06} recently
confirmed that the ISM density is indeed
low, $N_\mathrm{H} < 0.1$~cm$^{-3}$. The example of SN~1006 illustrates that
real difficulty in reliable prediction of expected \gr emission from SNRs is not
a theoretical one, but rather due to poorly known values of relevant astronomical
parameters, like distance to the object and ambient ISM density.
>From theoretical consideration \citep{kbv05} it becomes clear that
in the case of SN~1006 and in other similar cases \gr measurements will 
at the same time give a reliable estimate of the ambient ISM density.

Similar situation is for the two other cases of Cas~A and Tycho's SNR,
considered by \citet{hillas05}: renormalization of his estimates, made in spherical
approach, and the adoption the most reliable astronomical parameters provide the
value of the expected \gr flux consistent with the existing
measurements (or upper limits) \citep[see][]{ber05}. 

\section{Amplified magnetic field in SNRs}
All young SNRs (including three above cases) to which nonlinear kinetic
acceleration theory has been successfully applied up to now exhibit the strongly
amplified magnetic field as result of very efficient acceleration of nuclear
CRs \citep{vbk05,ber05}. 
Magnetic field extracted from the spectral properties of the overall
synchrotron emission and from the observed filamentary structure of the
nonthermal X-ray emission agree very well and their value can be expressed in
the form
\begin{equation}
B_0^2/(8\pi)=5\times 10^{-3}P_c,
\end{equation}
where $P_c$ is CR pressure at the
shock front, $B_0\gg B_{ISM}$ is the upstream field,
which is already amplified and therefore considerably exceeds magnetic field
value $B_{ISM}\approx 5$~$\mu$G in the ambient ISM. 
Let us consider the consistency of such a field strength with the
plasma physics.

Several attempts have been undertaken to model highly nonlinear turbulent
magnetic field, created by accelerated CRs upstream of the strong shock
\citep{belll01,vlad06,amato06}. Two aspects of this process are of significant
importance: the amount of magnetic field amplification and the extent of of gas
heating due to the turbulence dissipation. Unfortunately up to now it is not
clear in which proportions CR working on the gas in the precursor region is
divided between gas heating and magnetic field amplification. Therefore we
restrict ourselves here by a simple estimates in order to demonstrate that
considerable magnetic field amplification $B_0\gg B_{ISM}$, consistent with
Eq.(2) is indeed possible upstream of the strong shock.

There are two different mechanisms operating upstream of the fast super-Alfv\'enic
shock and leading to the amplification of the chaotic magnetic field. The first
one is related with the resonant interaction of CRs with the Alfv\'en waves, which
due to the CR anisotropy triggers an instability, followed by the intense Alfv\'en
wave excitation \citep{bell78,mkv82}. The other one, recently proposed by \cite{bell04}, is
non-resonant amplification of MHD perturbations due to Lorentz force, created by
CR current.

Integral effect of resonant Alfv\'en wave excitation can be written in the form
\citep{mkv82}
\begin{equation}
2u \frac{dE_B}{dx}=c_A\frac{dP_c}{dx},
\end{equation}
if we suggest that the CR working $c_AdPc/dx$ goes on the 
excitation of the Alfv\'en waves.
Here $E_B=(\delta B)^2/(8\pi)$ is magnetic component of the 
Alfv\'en wave energy density,
$c_A=B/\sqrt{4\pi \rho}$ is Alfv\'enic speed, $u$ is the plasma flow speed
directed along the $x$ axis
to the shock situated at the point $x=0$. 
For the purpose of estimates we suggest quasistationary condition and
neglect the gas speed variation $u(x)$ in the upstream region $x<0$ due to the
shock modification by CR backreaction, taking $u=V_s$.
Assuming that the far upstream field
$B_0=B_{ISM}$ coincides with the ISM field $B_{ISM}$ and performing 
the integration of the above equation we have magnetic field energy density 
at the shock front 
\begin{equation}
E_B/E_{B0}=M_{A0} P_c/(\rho u^2),
\end{equation}
where $M_{A0}=V_s/c_{A0}$ is Alfv\'enic shock Mach number, $E_{B0}=B_0^2/(8\pi)$ is the 
energy density of the far upstream field. Since the strong shock
provides efficient CR acceleration with $P_c/(\rho u^2)\sim 1$ the above
expression shows that Alfv\'en wave amplitude grows to a large value $\delta B\gg
B_0$ in the case of strong shock with Mach number $M_{A0}\gg 1$.

Resonant magnetic field amplification should be supplemented by two effects:
non-resonant instability and the Alfv\'en speed variation $c_A(B)$ across the
precursor due to the increase of the mean effective field $B$.
In this case \cite[see][for details]{pell06} instead of (4) we have
\begin{equation}
E_B/E_{B0}=M_{A0} (B_{nr}/B_0)P_c/(\rho u^2),
\end{equation}
where
\begin{equation}
B_{nr}=B_0\sqrt{\frac{3u^5}{\Phi c_{A0}^4c} \frac{P_c}{\rho u^2}}
\end{equation}
is the mean field amplified non-resonantly. Here $\Phi=\mbox{log}(p_{max}/m_pc)$, 
$p_{max}$ is the maximum CR momentum.

Substituting $\Phi=10$ in the expression (5) we have
\begin{equation}
E_B=3\times 10^{-2}(V_s/3000~\mbox{km/s})^{1/2}P_c.
\end{equation}
Since for all known young galactic SNRs the shock speed is about $V_s\approx
10^{-2}c$,
this expression 
shows, that existing mechanisms is able to amplify magnetic field within the
precursor to the level, which considerably exceeds the required value (2).
Therefore we can suggest that the main amount of the energy given by the
expression (5) in
fact goes on the gas heating due to dissipation of MHD perturbations and only
small fraction of it, given by (2), indeed goes on the 
magnetic field amplification.
Numerical simulation of considered process performed by \citet{bell04}
confirmed, that the gas thermal energy an order of magnitude larger than the
energy content of amplified magnetic field.

Note that the process of magnetic field amplification is 
not included in kinetic nonlinear theory \citep[e.g.][]{byk96,ber05}.
It simply postulates the existence of already amplified
far upstream field $B_0$, which value is consistent with experimental
requirements, given by Eq.(2). The consistency of such a picture is provided by
the fact, that
the spectrum of CRs produced by strong
modified shock is very hard so that CRs with highest energies have a largest
contribution in their energy content. These the most energetic CRs produce
field amplification on their spacial scale that is the precursor size.
Therefore CRs with lower energies already 'see' the amplified field $B_0\gg
B_{ISM}$. 
Note also, that kinetic nonlinear model contains in the explicit form the 
suggestion, that the whole 
CR working goes on the gas heating presumably due to some kind of dissipation
mechanism.
Since according to
the above estimates, the magnetic field energy which can be potentially achieved in the
precursor region is considerably larger than required energy, given by Eq.(2),
it can be considered as indirect evidence for the existence of such dissipation
processes, which efficiently converts the energy of magnetic field perturbations
into the gas thermal energy.

The existence of nonadiabatic gas heating within the precursor plays an
important role for the shock structure and dynamics. It restricts the shock
compression ratio by the value \citep{byk96,berel99}
\begin{equation}
\sigma\approx 1.5 M_{A0}^{3/8},
\end{equation}
if the whole CR working $c_AdP_c/dx$ goes on the gas heating. For the
appropriate upstream magnetic field values, consistent with Eq.(2), 
it provides moderate shock modification with $\sigma \approx 6$.
In the opposite case, when 
within the precursor there are no any additional gas heating except
adiabatic one, the expected shock compression ratio is \citep{byk96,berel99}
\begin{equation}
\sigma\approx 1.3 M_{S0}^{3/4},
\end{equation}
where $M_{S0}=V_s/c_{S0}$ is the sonic Mach number, $c_{S0}$ is the far upstream
sound speed. For the typical ISM parameters this expression leads to much larger
shock compression $\sigma\gg 6$.

In the case when CR working in the upstream region 
goes on the gas heating, the expected gas thermal
energy just ahead of the subshock is
\begin{equation}
E_{g}\approx P_c/M_{A0}.
\end{equation}
Appropriate magnetic field value $B_0$ given by Eq.(2) 
leads to a typical shock compression
ratio $\sigma=6$, that according to Eq.(8)
corresponds to $M_{A0}\approx 40$, that in turn according to Eq.(10) provides
gas thermal energy $E_{g}\approx
2.5\times 10^{-2}P_c$. This value $E_{g}$ very well corresponds to what is expected
from the expression (7),
if indeed main amount of magnetic field perturbation energy is dissipated.
We therefore conclude that kinetic nonlinear theory provides the consistent
description at least at shock speeds $V_s\sim 3000$~km/s, that correspond to the
most active phase of the SNR evolution that is the beginning of the Sedov phase.

\section{Maximum energy of CRs}
The most important consequence of magnetic field amplification in SNRs is the
substantial increase of the maximum energy of CRs
$\epsilon_{max}=p_{max}c$, accelerated by SN shocks. Let us briefly analyse what
value of $\epsilon_{max}$ can be expected taking into account the
amplification of SNRs magnetic field.

According to theoretical consideration \citep{lb00}
and the analysis of the observational
data \citep{pariz06} CR diffusion in the amplified magnetic field is close to
the Bohm limit. Therefore
cutoff momentum of CRs accelerated at any given SNR evolutionary stage 
can be expressed in the form
\citep{ber96}
\begin{equation}
p_m=m_pc R_sV_s/[A\kappa(m_pc)],
\end{equation}
where $\kappa(p)=pc^2/(3ZeB_0)$ is CR diffusion coefficient, corresponding to Bohm
limit; $Z$ is the charge number of CR nuclei. The value of parameter $A$ depends
on the shock expansion law $R_s(t)$ and on the downstream gas flow radial
profile $w(r)$. Note that the main factors which determines the value of CR
cutoff momentum $p_m$ are: i) finite shock size; ii) shock deceleration and iii)
CR adiabatic cooling in the expended downstream region $r<R_s$.
These factors are more relevant compared with the time factor, studied by \citet{lagc83},
that in turn prevents the complete shock modification due to CR backreaction
\citep{dvb95}.

First consider the most simple case of uniform ISM, that is typical for type Ia
SNe. In such a case the main part of CR spectrum formed during the SNR
evolution is produced at the Sedov phase.
Since the CR cutoff momentum at the Sedov phase
$p_m\propto R_sV_sB_0\propto R_sV_s^2\propto t^{-4/5}$ decreases with time $t$  
the most energetic CRs in the power law part of this spectrum
are produced at the very end of the free expansion phase. 
Therefore the maximum CR momentum 
$p_{max}=\mbox{max}\{p_m(t)\}$ depends on physical parameters as 
$p_{max}\propto V_0 R_0 B_0(t_0)$, where $V_0$, $R_0$ and $t_0$ are mean ejecta
speed, sweep up radius and sweep up time respectively. 
Taking into account the value of upstream
magnetic field (2) and assuming $P_c=0.5\rho_0 V_s^2$, we have \citep{bv04}
\begin{equation}
\frac{p_{max}}{m_pc}=
3\times 10^6 Z\left( \frac{E_{sn}}{10^{51}~\mbox{erg}}\right)
\left(\frac{M_{ej}}{1.4M_{\odot}}\right)^{-2/3}
\left(\frac{N_H}{0.3~\mbox{cm}^{-3}}\right)^{1/6}.
\label{pmax}
\end{equation}
It has a strong dependence on the SN parameters $E_{sn}$ and
$M_{ej}$, but is only weakly dependent upon the ISM density $\rho_0=1.4N_Hm_p$.

Maximum momentum of protons, given by Eq.(12) for $E_{sn}=10^{51}$~erg,
$M_{ej}=1.4M_{\odot}$ and $N_H=0.3$~cm$^{-3}$
is $p_{max}\approx 3\times 10^6m_pc$. This required 
condition to reproduce
the spectrum of GCRs up to the knee energy is fulfilled due
to the magnetic field amplification.

For CR iron nuclei $p_{max}\approx 10^8m_pc$ that provides the
possibility for the
formation of GCR spectrum up to the energy $\sim 10^{17}$~eV inside
SNRs. 

Since during the early free expansion SNR evolutionary phase the SN shock speed 
$V_s=(3-4)\times10^4$~km/s is
considerably higher than on the subsequent Sedov phase
($V_s=(3-10)\times10^3$~km/s), one could expect the
production of CRs with energies $p>p_{max}$ essentially higher than produced on
Sedov phase \citep{belll01}. Let us briefly consider this possibility, which was
studied by \citet{bv04} and by \citet{pz05}.

During the free expansion phase the SN shock evolution is determined by the 
velocity distribution of the fastest part of the ejecta which is described by a power law
\begin{equation}
dM_{ej}/dv\propto v^{2-k}
\label{Mej}
\end{equation}
with $k=7$~to~12 \citep{jss81,chev82}.
The ejecta kinetic energy contained in the fraction with speed
$v$ within the interval $dv$ is
$dE_{ej}(v)\propto v^{4-k}dv$.
The CR spectrum created by the strongly modified shock is quite
hard: $N(p)\propto p^{-\gamma}$, $\gamma <2$. For the sake of a rough
estimate we assume that all the above ejecta energy goes into CRs
with maximum momenta $p_m(V_s)$, where the shock speed is roughly the
ejecta speed $V_s\approx v$. This gives
\begin{equation}
N(p_m)p_mdp_m\propto dE_{ej}
\label{NEej}
\end{equation}
in the ultrarelativistic case.
Taking into account that $p_m\propto R_sV_sB\propto
V_s^{\alpha}$, we obtain 
\begin{equation}
\gamma=2+(k-5)/\alpha.
\label{gamma}
\end{equation}
Since in the case of uniform ISM $R_s\propto t^{(k-3)/k}$ we have $\alpha=(9-k)/3$
if we suggest for the simplicity $B_0\propto \rho_0V_s^2$. 
Even in the case of type Ia SNe with a typical value $k=7$ Eq.(15) gives
$\gamma=5$ \citep{bv04,pz05}. It means that the spectrum of CRs
with momenta $p>p_{max}$
accelerated during the free expansion phase is so steep, 
that it hardly plays any role for the formation of the
GCR spectrum.

In the case of type II/Ib SNe, which are more
numerous in our Galaxy \citep{tamm94}, during the initial SNR evolutionary phase
SN shock propagates through the progenitor star wind. Therefore in this case
ambient gas density at the shock front varies as $\rho_0\propto R_s^{-2}$ that
gives $\alpha=2$. Even at the lowest value $k=8$ the resulting CR spectrum is
very steep, $\gamma=3.5$.

We therefore conclude, that CRs accelerated during the free expansion phase do
not play a role for GCR spectrum, except the case if the actual value of
the parameter $k$ is as small as $k<5.4$ for type Ia SNe and $k<6$ for type
II/Ib SNe or the CSM has more complicated structure than considered above.

In Fig.8 we present GCR energy spectra, expected in the Galaxy, assuming that SNRs are the
main GCR source. Compared with our previous calculations \citep{bk99} much
higher magnetic field in SNR is used, which varies during the SNR evolution
according to Eq.(2). To obtain these spectra $J(\epsilon_k)\propto
J_s(\epsilon_k)\tau_{esc}(R)$ the corresponding source spectra
$J_s(\epsilon_k)$, produced in SNR during its evolution,
were calculated for a typical set of relevant
parameters ($E_{sn}=10^{51}$~erg, $M_{ej}=1.4M_{\odot}$,
$N_H=0.3$~cm$^{-3}$), taking into account their steepening due to the GCR escape
from the Galaxy with a time scale $\tau_{esc}\propto R^{-\mu}$,
where $R(\epsilon_k)$ is the rigidity, $\epsilon_k$ is the kinetic energy of CR
particle \citep{ber06}.

For a comparison with the experiment we represent in Fig.8 the data of only
three experiments, ATIC-2 \citep{atic2}, JACEE \citep{jacee}
and KASCADE \citep{kascade}, 
which on our view are 
representative in the corresponding energy ranges.
One can see, that the theory in a satisfactory way fits the existing data up to
the energy $\epsilon_k \approx 10^{17}$~eV. The essential exception is the
measured in the recent ATIC-2 balloon
experiment \citep{atic2} helium spectrum, which 
opposite to the theoretical expectation is noticeably flatter than the protons
spectrum.
Note that the spectral shape of protons and
helium measured in the previous ATIC-1 experiment 
at energies $10-10^4$~GeV \citep{atic06a} 
are very similar. Similar form of the spectra for different GCR nuclei at
energies $\epsilon_k=10^3-10^6$~GeV if observed could be considered as the
indirect confirmation of GCR origin in SNRs. Therefore the confirmation of
ATIC results for protons and helium by other experiments with comparable or
even better precision is very much needed.

Another difficulty for the theory is the fact that to get a consistency with the
observed GCR spectra the value $\mu=0.75$ is needed, which is a bit beyond the
experimentally determined interval $\mu=0.3-0.7$ \citep{berezinetal90}.

In addition, strong energy dependence of the GCR escape time
$\tau_{esc}(\epsilon)$ makes it difficult to interpret the observed CR
anisotropy, which is very low, less than one percent at energies
$\epsilon=10^{15}-10^{17}$~eV \citep{hillas05}. 
Here $\epsilon$ is total energy of CR particle.
Total flux of GCRs escaping from
the Galaxy $F_c\propto \epsilon^{-\gamma}$ has energy dependence like CR source
spectrum $J_s\propto \epsilon^{-\gamma}$ with $\gamma\approx2$. Therefore GCR
anisotropy outside the region where CR sources are situated (
Galactic disk volume) increases with energy $F_c/J\propto \epsilon^{\mu}$, like inverse
residence time $\tau_{esc}^{-1}$. If $\mu$ is as big as 0.6 the expected
anisotropy at $\epsilon =10^{17}$~eV exceeds the level of 100\%
\citep{hillas05}. Since our observing position is not outside the source layer,
we should see a smaller GCR anisotropy \citep{hillas05}. In addition one can
also suggest that in the region, occupied by CR sources, GCR diffusion
coefficient is higher than in outer part of GCR confinement volume, due to
enhanced level of turbulence produced by GCRs streaming away from the Galaxy.
In such a case the GCR anisotropy in the source layer could be sufficiently
small and consistent with the observations.

Note that the theory predicts concave spectra at energies $\epsilon>10Z$~GeV
for individual CR species and there are some experimental evidences that the
actual GCR spectra are indeed concave, eventhough one needs more precise
measurements at $\epsilon>10^4$~GeV in order to make more strict conclusion.
At the same time calculated all particle spectrum has almost pure power law 
form up to the knee energy (see Fig.8).

According to Fig.8  the knee
in the observed all particles
GCR spectrum has to be attributed to the maximum energy of
protons, produced in SNRs. The steepening of the all particle GCR spectrum above the knee
energy $3\times 10^{15}$~eV is a result of progressive depression of
contribution of light CR nuclei with the energy increase. Such a scenario 
is confirmed by KASCADE experiment which shows relatively sharp cutoff
spectra of GCR species at energies $\epsilon_{max}\approx 3Z\times10^{15}$~eV
\citep{kascade}. At energy $\epsilon \sim 10^{17}$~eV GCR spectrum is
dominated by iron nuclei contribution. 

According to the analysis, performed by  \citet{aloiso06} the contribution of
Galactic CR sources into the observed GCR spectrum dominates up to the energy
$10^{17}$~eV above which it progressively goes down so that at energy $10^{18}$~eV
the contribution of extragalactic sources becomes dominant.
At so-called second knee energy $\epsilon_k=5\times 10^{17}$~eV
these two sources provide roughly equal contributions into GCR spectrum.
As it is seen from Fig.8, calculations provide a good fit of the observed GCR up to the
required energy $\epsilon_k=10^{17}$~eV, that is a strong argument for SNRs to be
considered as the main Galactic CR source. 

Essentially different scenario of the formation of GCR spectrum and the first
knee in particular was proposed and discussed by 
Erlykin and Wolfandale in a series of papers \citep[see][and references there]{erlf00}.
Based on the analysis of measurements performed by ground based installations
they claim the existence of complex feature of GCR spectrum in the knee 
energy region. This leads them to a conclusion that observed GCR spectrum
consists of smooth background spectrum (without the
knee), produced by many CR sources, and superimposed single source 
(presumably nearby SNR) contribution, that forms a bump on the smooth background
spectrum usually interpreted as a (first) knee.
The contribution of the single nearby SNR, leading
to some peculiarity of GCR spectrum at the detectable level
in the knee region, is indeed quite possible \citep{bk99}.
However, the existing measurements of CR composition 
demonstrate, that the knee is indeed peculiar point in GCR spectrum, where not
only the shape of GCR spectrum but also their chemical composition undergoes 
a noticeable change, that is consistent with CR production in typical SNRs, as
described above. 

\section{Summary}
Detailed consideration performed within a frame of nonlinear
kinetic theory demonstrates
that the CR production efficiency in young SNRs is consistent with requirements
for the Galactic CR sources. Recent experimental and theoretical studies of
Tycho's SNR, \rxj and SN~1987A provides new very strong evidence for that.

Synchrotron radiation from young SNRs provides evidence that efficient CR
acceleration followed by the significant shock modification and
magnetic field amplification takes place there.
Magnetic field amplification leads to the considerable increase of maximum
energy of CRs, accelerated in SNRs. Calculations performed within kinetic
nonlinear theory demonstrate that the expected GCR spectrum, produced in SNRs,
in satisfactory way fits the existing GCR data up to the energy $10^{17}$~eV. 
The first knee
in the observed all particles
GCR spectrum is attributed to the maximum energy of
protons, produced in SNRs. The steepening of the all particle GCR spectrum above the knee
energy $3\times 10^{15}$~eV is a result of progressive depression of
contribution of light CR nuclei with the energy increase. Such a scenario 
is confirmed by KASCADE experiment which shows relatively sharp cutoff
spectra of GCR species at energies $\epsilon_{max}\approx 3Z\times10^{15}$~eV
\citep{kascade}.

The nature of the whole GCR spectrum, which does not require any
significant contribution of galactic CR sources other then SNRs, can be the
following. GCR spectrum up to the energy $\epsilon=10^{17}$~eV is dominated by the
contribution of galactic sources, whereas at $\epsilon>10^{18}$~eV GCRs are
predominantly extragalactic. Such a picture is consistent with the analysis,
presented here, and with the analysis of ultra high energy CR properties
\citep{aloiso06} as well. Detail experimental study of GCR spectrum and
composition could find whether indeed the transition from galactic to
extragalactic component takes place in energy range $10^{17}-10^{18}$~eV.

Due to high interior magnetic field in young SNRs the 
$\pi^0$-decay $\gamma$-rays generated by the
nuclear CR component as a rule dominates over $\gamma$-rays, generated by
electron CR component, and calculated $\gamma$-ray flux fits existing data.

New generation of ground-based $\gamma$-ray detectors CANGAROO~III, H.E.S.S., 
MAGIC, VERITAS will presumably 
measure significantly more accurate $\gamma$-ray spectra from 
already detected SNRs, like Cas~A, RXJ1713.7 and Tycho's SNR, as well as
from a number of new sources. 
In this regard considerable interest as a potential \gr sources represent
Tycho's SNR, SN~1987A and Kepler's SNR \citep{bkv06} eventhough the expected \gr
fluxes from these objects only slightly above the sensitivity of modern TeV \gr
instruments.
Together with data, which soon will be obtained by space
$\gamma$-ray observatory GLAST, they will reproduce the full 
$\gamma$-ray spectra of some
number of individual SNRs, as well as the
Galactic background spectrum. 
Although the existing data
together with their theoretical interpretation presents a consistent picture 
of GCR origin,
these new measurements will allow for detailed study of SNRs and the CR production
inside them. 

\section{Acknowledgments}
This work has been supported by the Russian Foundation For Rasic Research
(grants 07-02-00221, 06-02-96008, 05-02-16412).
The author thanks the organizers of the COSPAR meeting E1.4 "New High-Energy
Results on
Supernova Remnants and Pulsar Wind Nebula" 
for the invitation to present this paper and for the financial support,
Prof. M.I.Panasyuk and Dr.A.D.Panov for their kind presentation of ATIC-2 data
and Prof. H.J.V\"olk for many useful discussions.

\pagebreak

%
%
%
\begin{figure}[t]
\centering
\includegraphics[width=7.5cm]{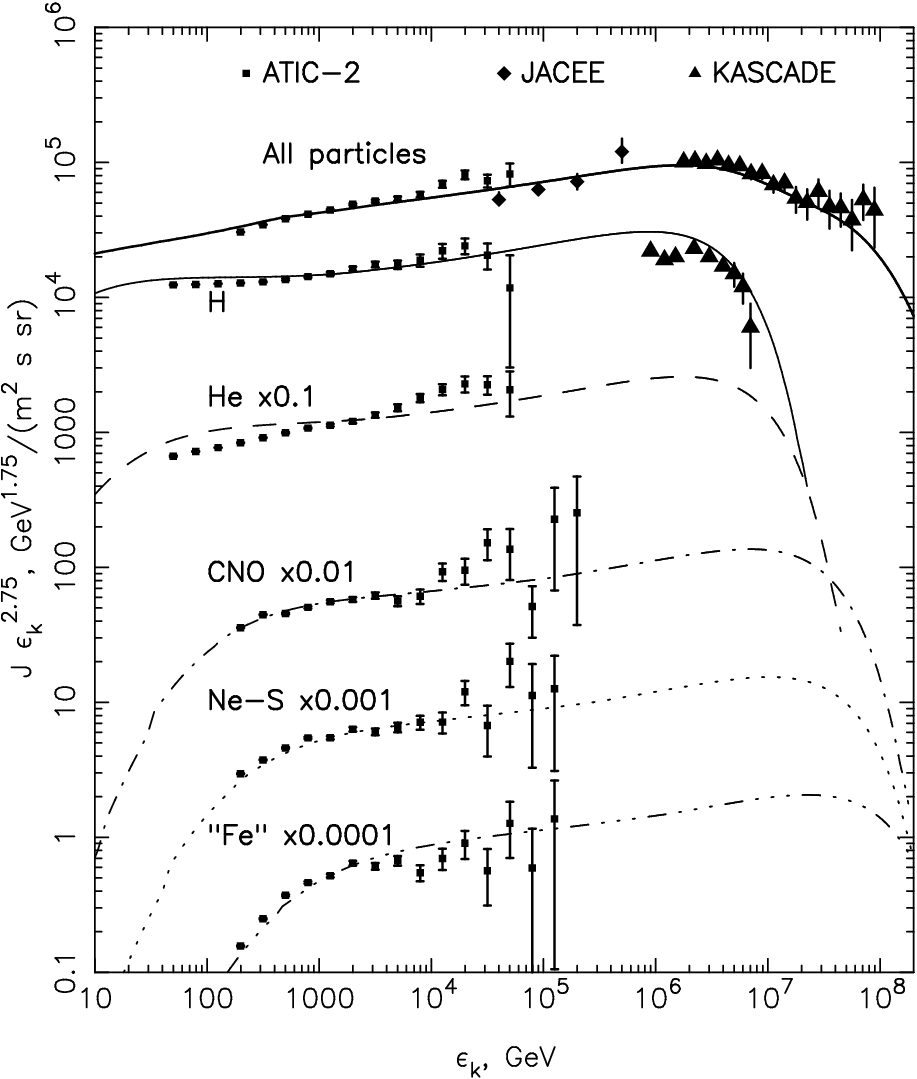}
\caption{Energy spectra for five groups of abundant GCR nuclei and all 
particles
spectrum \citep{ber06}. The data of ATIC-2 \citep{atic2}, JACEE \citep{jacee} and KASCADE
\citep{kascade} experiments are shown as well.
}
\label{f8}
\end{figure}

\end{document}